\RequirePackage{ifpdf}
\documentclass{pCHEF}

\title{The Time Structure of Hadronic Showers in Imaging Calorimeters with Scintillator and RPC Readout}

\ShortTitle{Time Structure of Hadronic Showers in Imaging Calorimeters}

\author{\speaker{Frank Simon}, on behalf of the CALICE Collaboration\\
        Max-Planck-Institut f\"ur Physik, Munich, Germany\\
        E-mail: \email{fsimon@mpp.mpg.de}}

\abstract{The intrinsic time structure of hadronic showers has been studied to evaluate its influence on the timing capability and on the required integration time of highly granular hadronic calorimeters in future collider experiments. The experiments have been carried with systems of 15 detector cells, using both scintillator tiles with SiPM readout and RPCs, read out with fast digitizers and deep buffers. These were installed behind the CALICE scintillator - Tungsten and RPC - Tungsten calorimeters as well as behind the CALICE semi-digital RPC - Steel calorimeter during test beam periods at the CERN SPS.  We will discuss the technical aspects of these systems, and present results on the measurement of the time structure of hadronic showers in steel and tungsten calorimeters. These  are compared to GEANT4 simulations, providing important information for the validation and the improvement of the physics models. In addition, a comparison of the observed time structure with scintillator and RPC active elements will be presented, which provides insight into the differences in sensitivity to certain aspects of hadronic showers depending on readout technology.}

\FullConference{Calorimetry for High Energy Frontiers - CHEF 2013,\\
		April 22-25, 2013\\
		Paris, France}

\usepackage{CHEF2013_definitions}

\begin{document}

\section{Introduction}

The time structure of hadronic showers plays a key role when evaluating the timing capabilities of calorimeter systems. This is of particular interest in the context of the development of detector concepts for the Compact Linear Collider (CLIC) \cite{Lebrun:2012hj}, where time stamping of signals on the nanosecond level is of key importance to reject pile-up from hadrons produced in two-photon processes, and where tungsten is used as absorber material for the hadronic calorimeters \cite{Linssen:2012hp}. This choice of a heavy absorber for the hadron calorimeter is expected to result in a particularly rich time structure of the hadronic cascade. The sensitivity to neutrons in the later parts of the hadronic cascade also influences the spatial structure of the visible signal in the detector, and is thus of importance for the performance of particle flow algorithms \cite{Thomson:2009rp, Marshall:2012ry} . These algorithms, which are used for jet reconstruction with unprecedented precision in linear collider detectors, rely on two-particle separation in the calorimeters. The spatially resolved measurement of the time structure of showers and the comparison of detection media with different sensitivity to neutrons, such as plastic scintillators and gaseous detectors, is thus of high relevance for the development of calorimeter technologies for such a future collider. The comparison of the measurements to simulations provide a means of validating the modelling of the time structure of the showers by different GEANT4 \cite{Agostinelli:2002hh} physics lists. Due to limited previous experimental input and the expected larger effects of late shower components in tungsten-based calorimeters this is particularly relevant for the case of tungsten absorbers.

\section{Experimental Setup}

\begin{figure}
\centering
\includegraphics[height=0.35\textwidth]{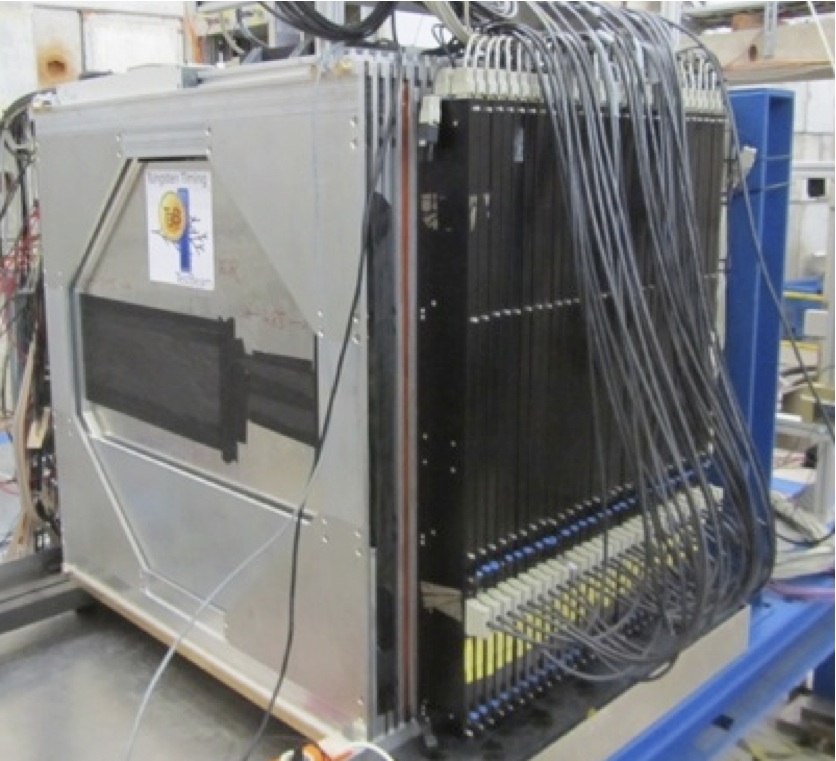}
\hspace{0.5 cm}
\includegraphics[height=0.35\textwidth]{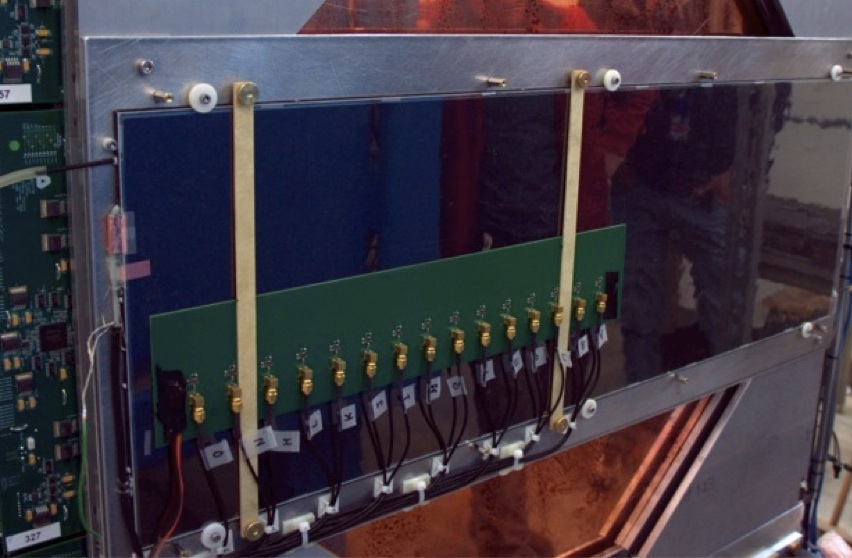}
\caption{The T3B setup downstream of the CALICE analog tungsten HCAL (left) and the FastRPC setup downstream of the CALICE digital tungsten HCAL, showing the RPC and the readout board mounted to it (right).}
\label{fig:T3BInstallation}
\end{figure}

The average time structure of hadronic showers is measured by the T3B (Tungsten Timing Test Beam) and the FastRPC experiments,
which were taking data together with the CALICE analog scintillator tungsten HCAL (WAHCAL) \cite{LucaciTimoce201388, Adloff:2010hb} and the CALICE digital RPC tungsten HCAL (WDHCAL) \cite{Bilki:2008df}, with the active elements of both calorimeters which were originally operated with steel absorbers using the same tungsten absorber structure in the CERN test beams.  In addition, the T3B experiment also took data with CALICE semi-digital HCAL (SDHCAL) \cite{Laktineh:2011zz} with steel absorbers to provide a comparison of the impact of different absorber material. The T3B  setup \cite{Soldner:2011np} consists of 15 scintillator cells with a size of $3\,\times\,3$ cm$^2$ and a thickness of 5 mm, with directly coupled Hamamatsu MPPCs following the scintillator tile design presented in \cite{Simon:2010hf}, while FastRPC uses a glass RPC \cite{Drake:2007zz} with pad readout with a geometry identical to that of T3B. The 15 cells / pads are arranged in one row extending from the center of the
calorimeter out to one side of the detector, covering the full radial extent of the showers. Figure \ref{fig:T3BInstallation} shows the T3B and the FastRPC setups installed downstream of the CALICE WAHCAL and WDHCAL, respectively.  

Both systems use the same readout chain, starting with a custom-designed pre-amplifier board which feeds the analog signals from the SiPMs and the RPC into a set of four 4-channel USB-oscilloscopes\footnote{PicoTech PicoScope 6403 (http://www.picotech.com/)} which provide a sampling rate of 1.25\,GSa/s on all channels and are therefore well suited for precise timing measurements in the nanosecond region. Long acquisition windows of \mbox{2.4 $\mu$s} per event are recorded to study the time structure of the energy deposits in the active medium in detail, providing information on the time structure of hadronic showers in the calorimeter.
 
The small number of channels  is insufficient for event-by-event measurements, but is used to measure the average time structure of showers in large data samples. In addition, the information from the main calorimeters can be used to reconstruct the  position
of the first inelastic hadronic interaction event by event, allowing to measure the time structure of the shower at various depths with
respect to the shower start, which can be used to measure the averaged timing profile over the full longitudinal and lateral extent of the shower.

In addition to the measurements performed with tungsten absorbers, T3B also took data together with the CALICE semi-digital calorimeter with steel absorbers to provide a comparison of the two absorbers.

\section{Results - The Time Structure in Tungsten with Scintillators and RPCs}

\begin{figure}
\centering
\includegraphics[width=0.75\textwidth]{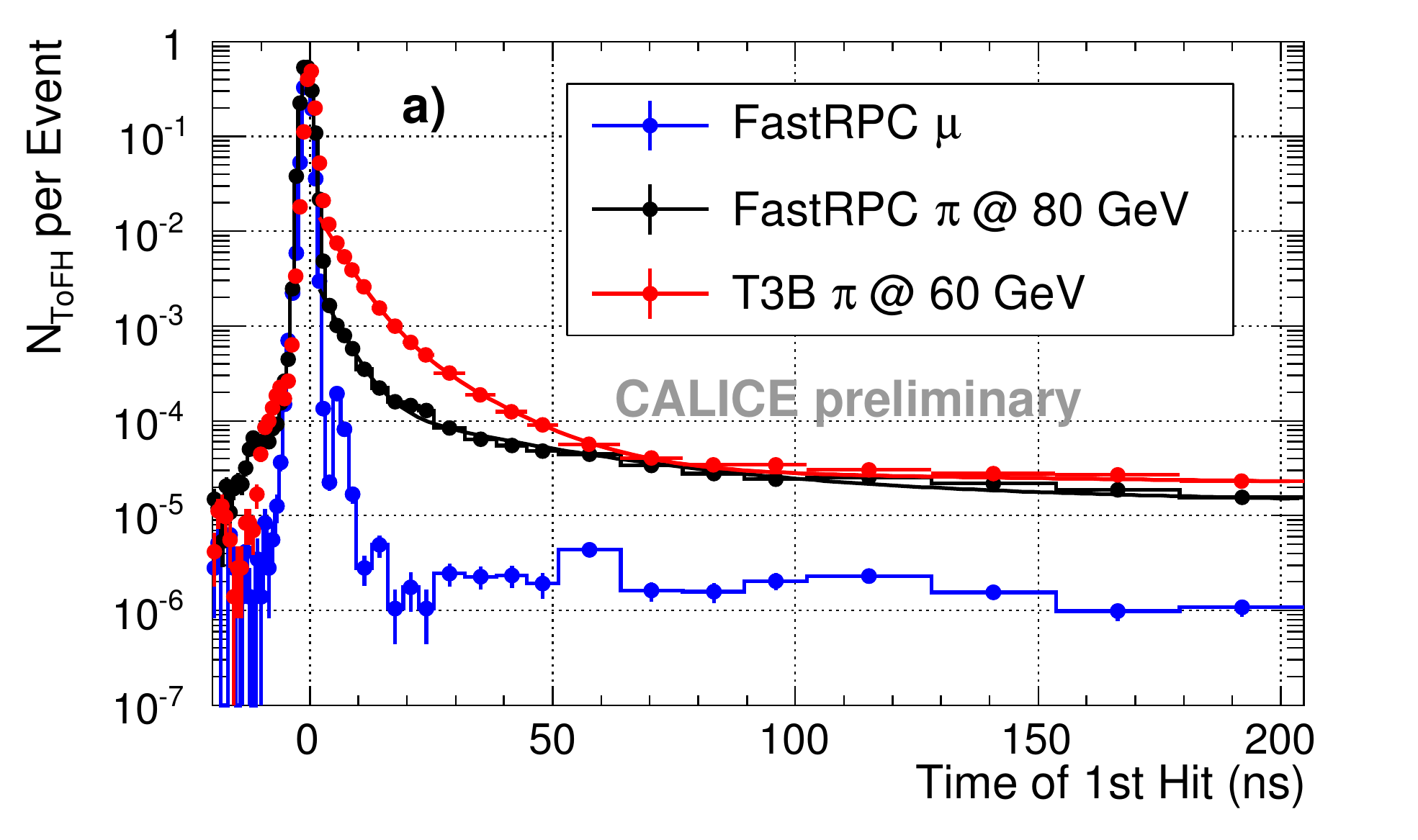}
\caption{Comparison of the time of first hit with scintillator and RPC readout. For reference, the time distribution for muons in FastRPC is also shown.}
\label{fig:ScintRPC}
\end{figure}

For both experiments, a sophisticated calibration and reconstruction framework has been developed. In the case of scintillator readout, this system is capable of determining the arrival time of each photon on the photon sensor on the nanosecond level by iteratively subtracting single photon signals from the recorded waveform. Further analysis is then performed on the photon time distribution. With RPC readout, the recorded pulse is used directly in the analysis. To provide a robust base for comparison between the two systems and to eliminate effects from afterpulsing of the SiPMs, the time of first hit is studied, which is defined by the time of the first energy deposit corresponding to at least the equivalent of 0.3 minimum-ionizing particles within 9.6 ns in a given cell in an event. Due to the high granularity of the readout, the probability for multiple hits in one cell in one event is on the percent level. The use of the time of first hit rather than using all observed hits thus does not result in an appreciable bias. 

Figure \ref{fig:ScintRPC} shows the distribution of the time of first hit, normalized to the number of events with at least one hit in the timing layer for both scintillator and RPC readout. This normalization also accounts for the difference in beam energy, which was found to not result in significant differences in this distribution. Here, the data sets with the highest statistics for the two experiments are used. For reference, the distribution for muons with RPC readout is also shown, indicating the distribution observed with an instantaneous signal. It is clearly apparent that hadronic showers lead to a substantial late signal component with both readout types that extends substantially beyond \mbox{200 ns}. The figure also shows a discrepancy of up to a factor of eight in the time region from 10 ns to 50 ns, a region where signals are expected to originate to a large extent from MeV-scale spallation neutrons. Due to the low density and the low hydrogen content in the RPCs, their sensitivity to this component is significantly reduced compared to plastic scintillator.

\begin{figure}
\centering
\includegraphics[width=0.75\textwidth]{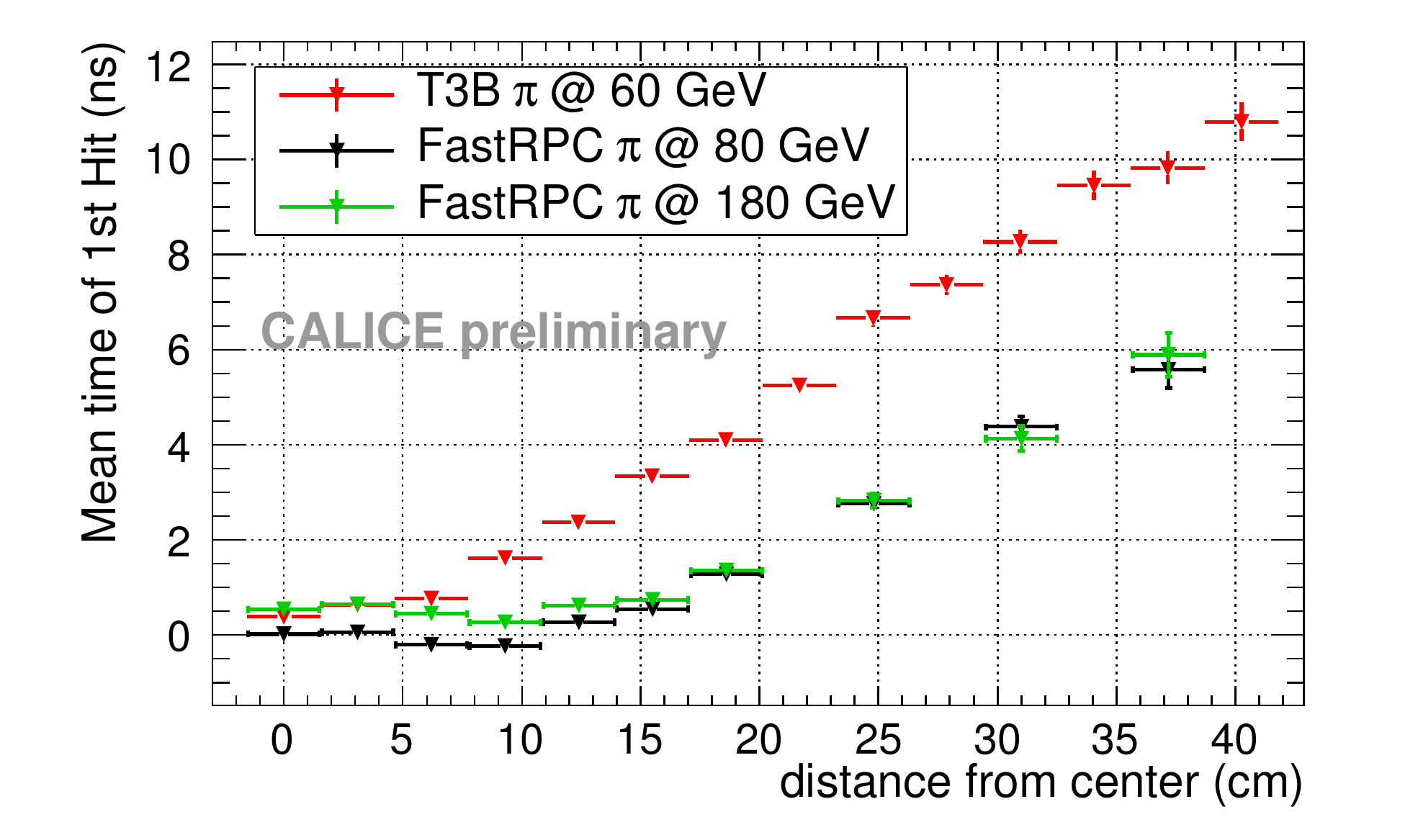}
\caption{Mean time of first hit in hadronic showers as a function of radial distance from the beam axis for scintillator and RPC readout.}
\label{fig:Radial}
\end{figure}

Figure \ref{fig:Radial} shows the radial distribution of the mean time of first hit, calculated up to 200 ns, again comparing gaseous and plastic scintillator readout. On the shower axis, the mean is close to zero in both cases due to the dominance of energy deposits from compact electromagnetic subshowers and from relativistic hadrons. The late components, which spreads out more widely than the relativistic particles due to the diffusion of neutrons, gain in importance at larger radii. Since the sensitivity in particular to the intermediate component is suppressed with for RPC readout the mean stays low out to larger radii compared to scintillator readout, making the core of the shower more compact in time with gaseous readout.

\section{Results - Comparison of T3B Data with GEANT4 Simulations}

\begin{figure}
\centering
\includegraphics[width=0.95\textwidth]{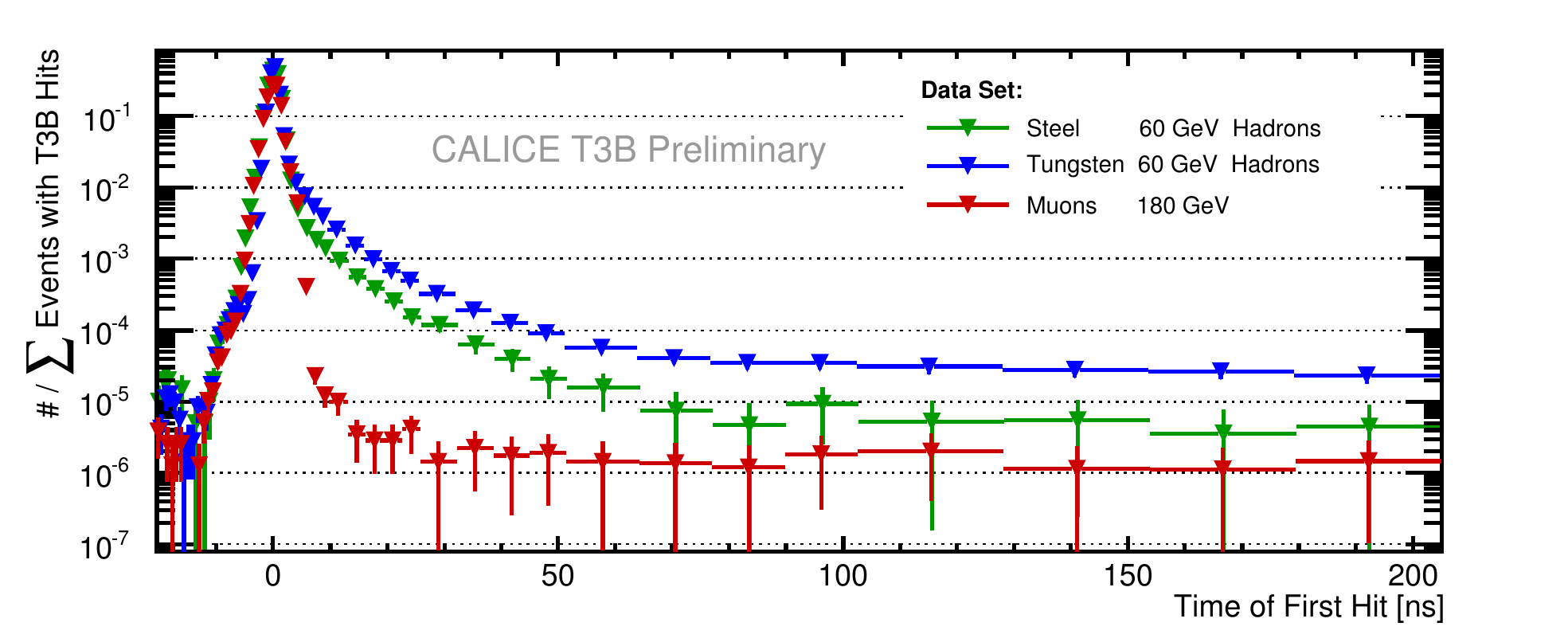}
\caption{Comparison of the time of first hit for muons (red) and 60 GeV pions in steel (green) and tungsten (blue) with the T3B plastic scintillator readout.}
\label{fig:TungstenvsSteel}
\end{figure}

For T3B, a sophisticated GEANT4-based simulation framework has been developed, which includes a detailed modelling of the detector response including detector effects on the time distribution of the signals. For FastRPC, a comparable simulation framework is not yet available, so we only discuss scintillator results here. Also, although the main emphasis of the T3B and Fast\-RPC programs is on the investigation of the time structure in a hadron calorimeter with tungsten absorbers, reference data with steel were also taken with scintillator readout. Figure \ref{fig:TungstenvsSteel} shows the comparison of the distribution of the time of first hit in tungsten and steel. In both cases, hadronic showers lead to a considerable late signal component compared to the muon reference. These late components are substantially more pronounced in tungsten than in steel, stressing the importance of a realistic modelling of the time structure when developing tungsten-based calorimeter systems for future collider detectors. 

\begin{figure}
\centering
\includegraphics[width=0.495\textwidth]{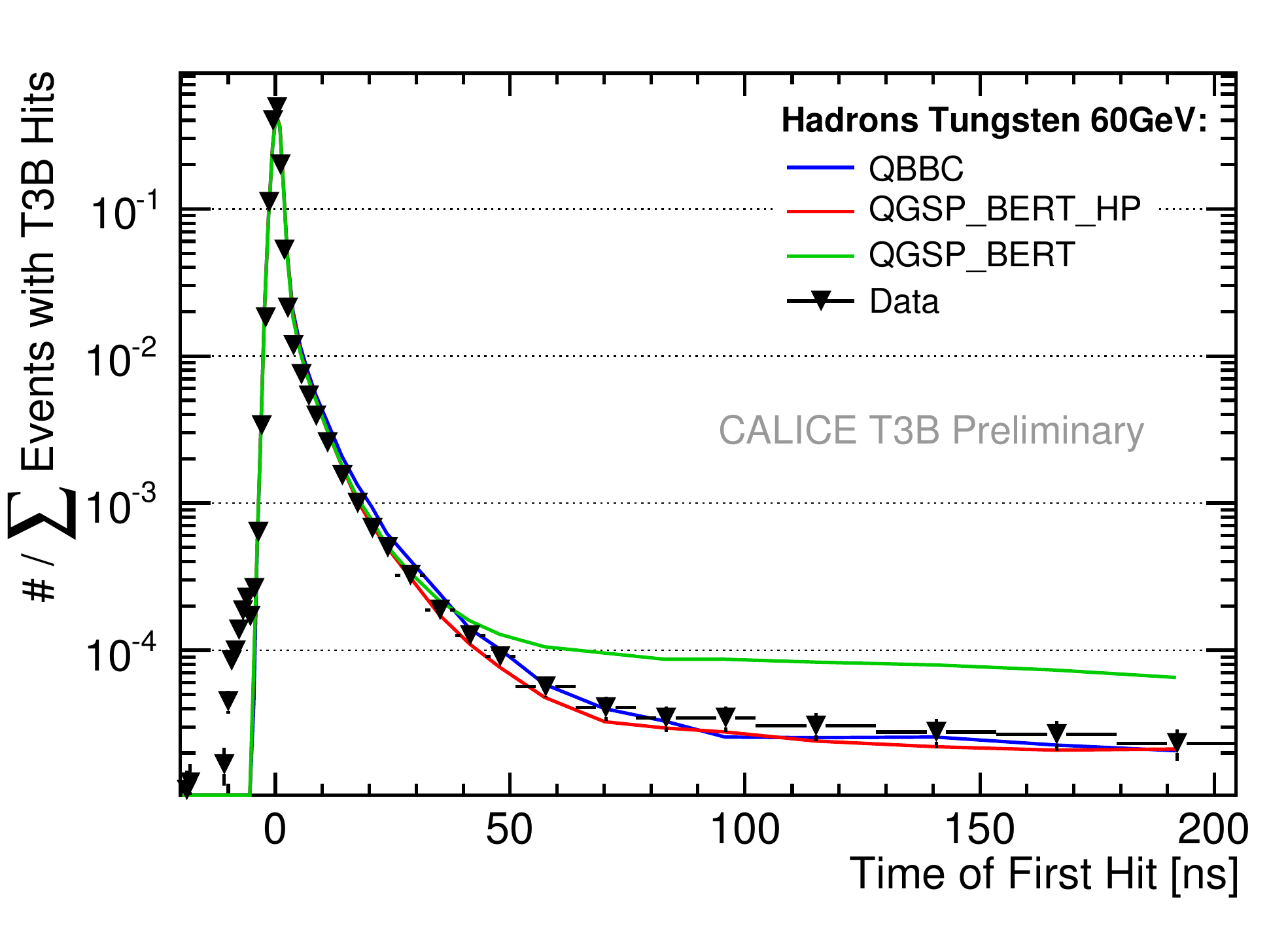}
\hfill
\includegraphics[width=0.495\textwidth]{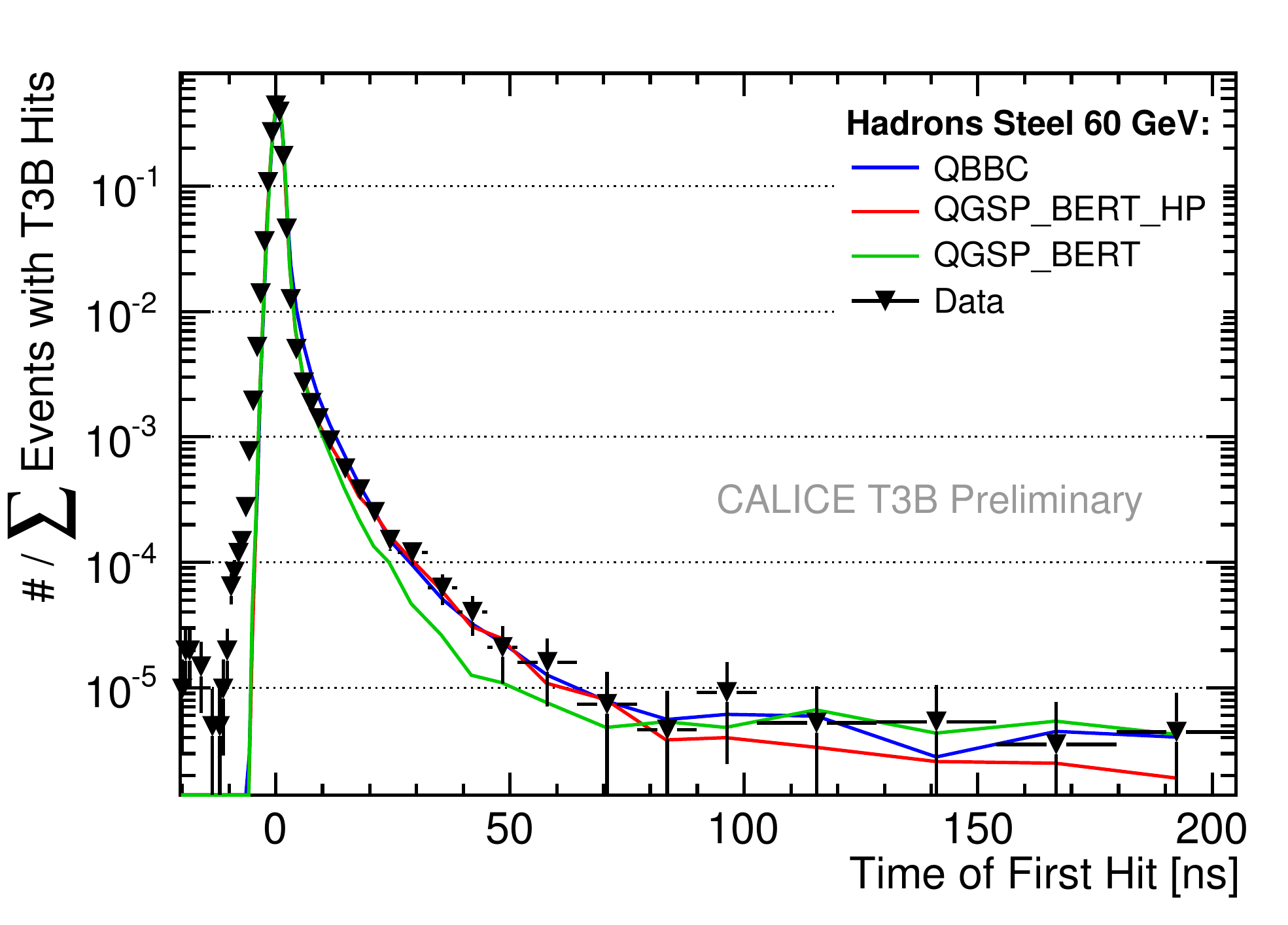}
\caption{Mean time of first hit in hadronic showers in tungsten (left) and steel (right) compared to GEANT4 simulations with three different physics lists. QGSP\_BERT\_HP and QBBC use specific models for low-energy neutrons.}
\label{fig:DataMCTimeDistribution}
\end{figure}

Figure \ref{fig:DataMCTimeDistribution} shows the distribution of the time of first hit observed in tungsten and steel compared to simulations with different physics models, which differ in particular in their treatment of low-energy neutrons  \cite{Geant4:PhysicsLists}. While  QGSP\_BERT\_HP and QBBC, which both have specialized neutron components, reproduce the observations in both tungsten and steel, QGSP\_BERT, which was the main production physics list for the LHC experiments and for the ILC and CLIC detector optimization studies, is only capable of describing the time structure in steel, while predicting significantly too much late shower activity in tungsten.

\begin{figure}
\centering
\includegraphics[width=0.7\textwidth]{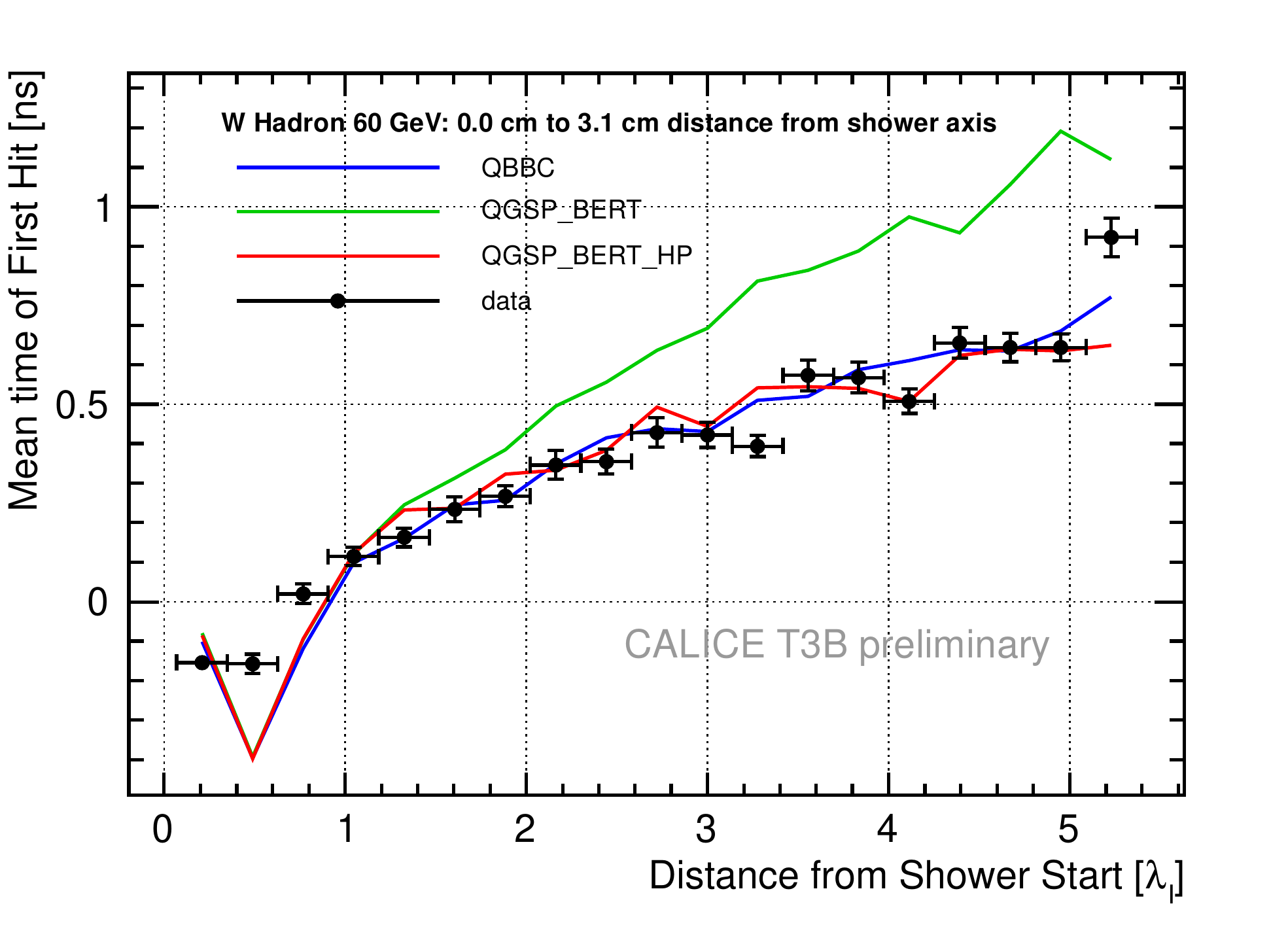}
\caption{Mean time of first hit along the beam axis measured in the central tile of T3B as a function of depth in the hadronic shower, compared to Geant4 simulations with different physics lists.}
\label{fig:LongitudinalProfile}
\end{figure}

During data taking with the WAHCAL, the T3B data acquisition was synchronized with the main CALICE DAQ, which allows a combined analysis of the data in tungsten. Due to the low trigger rate of the first prototype run of the complete SDHCAL, such a synchronization was not practical for the case of steel. The combined analysis provides the possibility to determine the point of the first inelastic interaction ("shower start") on an event-by-event basis. This provides a measurement of the depth of the T3B layer within the shower, which is used to reconstruct an average longitudinal time profile of hadronic showers in tungsten by combining the data of many events, similar to the radial time profile discussed above, which are by design averaged over all shower starting positions. Figure \ref{fig:LongitudinalProfile} shows the longitudinal profile of the mean time of first hit in the shower core along the shower axis compared to the three GEANT4 physics lists also used above. The data are by construction corrected for the time-of-flight of a ultra-relativistic particle since the T3B layer always has the same distance with respect to the trigger scintillators which set the event time. The measurement shows the dominance of prompt shower components in the front part of the cascade, driven by the instantaneous electromagnetic component and by relativistic hadrons, and some contributions of later energy deposits which gain in importance towards the rear of the shower. Only the models with special treatment of low-energetic neutrons are capable of reproducing the time structure of the deeper part of the shower, consistent with the observations made for the overall time distribution of the energy deposits discussed above.

\section{Summary and Conclusions}

With the two add-on experiments T3B and FastRPC, which were taking data together with the main hadron calorimeter prototypes, CALICE has extended its program to a spatially resolved study of the time structure of hadronic showers. The comparison of the observed time structure with plastic scintillator and with RPC readout in calorimeters with tungsten absorbers shows a substantially reduced sensitivity of the gaseous detectors to the intermediate time component from approximately 10 ns to 50 ns. This is consistent with the expectation that MeV-scale neutrons contribute strongly in this region, which are efficiently detected with plastic scintillator due to its high hydrogen content. This also results in differences in the radial timing profile, which is more dominated by prompt components in calorimeters with RPC readout. The comparison of the time structure observed with scintillators in a steel calorimeter with that observed with tungsten absorbers shows substantially increased late activity in the case of tungsten. While the time distributions in steel are generally well modelled by the standard QGSP\_BERT physics model in GEANT4, the reproduction of the tungsten results requires physics lists with high precision neutron treatment, such as  QGSP\_BERT\_HP, further demonstrating the increased influence of low-energy neutrons in calorimeters with heavy absorber material.

\bibliography{CALICE}{}

\begin{thebibliography}{10}

\bibitem{Lebrun:2012hj}
P.~Lebrun, L.~Linssen, A.~Lucaci-Timoce, D.~Schulte, F.~Simon, et~al.
\newblock {The CLIC Programme: Towards a Staged e+e- Linear Collider Exploring
  the Terascale : CLIC Conceptual Design Report}.
\newblock {\em arXiv:1209.2543 [physics.ins-det]}, 2012.

\bibitem{Linssen:2012hp}
Lucie Linssen, Akiya Miyamoto, Marcel Stanitzki, Harry Weerts, and Daniel~J.
  Feldman.
\newblock {Physics and Detectors at CLIC: CLIC Conceptual Design Report}.
\newblock {\em arXiv:1202.5940 [physics.ins-det]}, 2012.

\bibitem{Thomson:2009rp}
M.A. Thomson.
\newblock {Particle Flow Calorimetry and the PandoraPFA Algorithm}.
\newblock {\em Nucl.Instrum.Meth.}, A611:25--40, 2009.

\bibitem{Marshall:2012ry}
J.S. Marshall, A.~Münnich, and M.A. Thomson.
\newblock {Performance of Particle Flow Calorimetry at CLIC}.
\newblock {\em Nucl.Instrum.Meth.}, A700:153--162, 2013.

\bibitem{Agostinelli:2002hh}
S.~Agostinelli et~al.
\newblock {GEANT4: A Simulation toolkit}.
\newblock {\em Nucl. Instrum. Meth.}, A506:250--303, 2003.

\bibitem{LucaciTimoce201388}
A.~Lucaci-Timoce.
\newblock {Shower development of particles with momenta below 10 GeV in a
  highly granular scintillator-tungsten hadron calorimeter}.
\newblock {\em Nucl. Instrum. Meth.}, A718:88 -- 90, 2013.

\bibitem{Adloff:2010hb}
C.~Adloff et~al.
\newblock {Construction and Commissioning of the CALICE Analog Hadron
  Calorimeter Prototype}.
\newblock {\em JINST}, 5:P05004, 2010.

\bibitem{Bilki:2008df}
Burak Bilki, John Butler, Tim Cundiff, Gary Drake, William Haberichter, et~al.
\newblock {Calibration of a digital hadron calorimeter with muons}.
\newblock {\em JINST}, 3:P05001, 2008.

\bibitem{Laktineh:2011zz}
I.~Laktineh.
\newblock {Construction of a technological semi-digital hadronic calorimeter
  using GRPC}.
\newblock {\em J. Phys. Conf. Ser.}, 293:012077, 2011.

\bibitem{Soldner:2011np}
Christian Soldner.
\newblock {Scintillators with Silicon Photomultiplier Readout for Timing
  Measurements in Hadronic Showers}.
\newblock {\em IEEE Nucl.Sci.Symp.Conf.Rec.}, 2011:2060--2062, 2011.

\bibitem{Simon:2010hf}
F.~Simon and C.~Soldner.
\newblock {Uniformity Studies of Scintillator Tiles directly coupled to SiPMs
  for Imaging Calorimetry}.
\newblock {\em Nucl.Instrum.Meth.}, A620:196--201, 2010.

\bibitem{Drake:2007zz}
Gary Drake, Jose Repond, David~G. Underwood, and Lei Xia.
\newblock {Resistive Plate Chambers for hadron calorimetry: Tests with analog
  readout}.
\newblock {\em Nucl.Instrum.Meth.}, A578:88--97, 2007.

\bibitem{Geant4:PhysicsLists}
A.~Ribon, J.~Apostolakis, A.~Dotti, G.~Folger, V.~Ivanchenko, M.~Kosov,
  V.~Uzhinsky, and D.~H. Wright.
\newblock {Status of Geant4 hadronic physics for the simulation of LHC
  experiments at the start of LHC physics program}.
\newblock {\em CERN-LCGAPP-2010-02}, 2010.

\end{thebibliography}

\bibliographystyle{unsrt}

\end{document}